\documentclass[ALICE,manyauthors]{cernphprep}

\usepackage[comma,square,numbers,sort&compress]{natbib}
\usepackage{hyperref}
\usepackage{lineno}
\usepackage{color}
\usepackage{floatrow}
\usepackage{verbatim}
\usepackage[T1]{fontenc}
\usepackage{orcidlink}

\begin{document}

%

\newcommand{\pp}           {pp\xspace}
\newcommand{\ppbar}        {\mbox{$\mathrm {p\overline{p}}$}\xspace}
\newcommand{\XeXe}         {\mbox{Xe--Xe}\xspace}
\newcommand{\PbPb}         {\mbox{Pb--Pb}\xspace}
\newcommand{\pPb}          {\mbox{p--Pb}\xspace}
\newcommand{\AuAu}         {\mbox{Au--Au}\xspace}
\newcommand{\dAu}          {\mbox{d--Au}\xspace}

\newcommand{\roots}        {\ensuremath{\sqrt{s}}\xspace}
\newcommand{\snn}          {\ensuremath{\sqrt{s_{\mathrm{NN}}}}\xspace}
\newcommand{\pt}           {\ensuremath{p_{\rm T}}\xspace}
\newcommand{\meanpt}       {$\langle p_{\mathrm{T}}\rangle$\xspace}
\newcommand{\ycms}         {\ensuremath{y_{\rm CMS}}\xspace}
\newcommand{\ylab}         {\ensuremath{y_{\rm lab}}\xspace}
\newcommand{\etarange}[1]  {\mbox{$\left | \eta \right |~<~#1$}}
\newcommand{\yrange}[1]    {\mbox{$\left | y \right |~<~#1$}}
\newcommand{\dndy}         {\ensuremath{\mathrm{d}N_\mathrm{ch}/\mathrm{d}y}\xspace}
\newcommand{\dndeta}       {\ensuremath{\mathrm{d}N_\mathrm{ch}/\mathrm{d}\eta}\xspace}
\newcommand{\avdndeta}     {\ensuremath{\langle\dndeta\rangle}\xspace}
\newcommand{\dNdy}         {\ensuremath{\mathrm{d}N_\mathrm{ch}/\mathrm{d}y}\xspace}
\newcommand{\Npart}        {\ensuremath{N_\mathrm{part}}\xspace}
\newcommand{\Ncoll}        {\ensuremath{N_\mathrm{coll}}\xspace}
\newcommand{\dEdx}         {\ensuremath{\textrm{d}E/\textrm{d}x}\xspace}
\newcommand{\RpPb}         {\ensuremath{R_{\rm pPb}}\xspace}
\newcommand{\kstar}        {\ensuremath{k^*}\xspace}
\newcommand{\mt}           {\ensuremath{m_{\rm{T}}}\xspace}

\newcommand{\nineH}        {$\sqrt{s}~=~0.9$~Te\kern-.1emV\xspace}
\newcommand{\seven}        {$\sqrt{s}~=~7$~Te\kern-.1emV\xspace}
\newcommand{\thirteen}     {$\sqrt{s}~=~13$~Te\kern-.1emV\xspace}
\newcommand{\forteen}     {$\sqrt{s}~=~14$~Te\kern-.1emV\xspace}
\newcommand{\twoH}         {$\sqrt{s}~=~0.2$~Te\kern-.1emV\xspace}
\newcommand{\twosevensix}  {$\sqrt{s}~=~2.76$~Te\kern-.1emV\xspace}
\newcommand{\five}         {$\sqrt{s}~=~5.02$~Te\kern-.1emV\xspace}
\newcommand{\fivefive}     {$\sqrt{s}~=~5.5$~Te\kern-.1emV\xspace}
\newcommand{\twosevensixnn}{$\sqrt{s_{\mathrm{NN}}}~=~2.76$~Te\kern-.1emV\xspace}
\newcommand{\fivenn}       {$\sqrt{s_{\mathrm{NN}}}~=~5.02$~Te\kern-.1emV\xspace}
\newcommand{\LT}           {L{\'e}vy-Tsallis\xspace}
\newcommand{\tev}          {TeV\xspace}
\newcommand{\gev}          {GeV\xspace}
\newcommand{\mev}          {MeV\xspace}
\newcommand{\GeVmass}      {GeV/$c^2$\xspace}
\newcommand{\MeVmass}      {MeV/$c^2$\xspace}
\newcommand{\lumi}         {\ensuremath{\mathcal{L}}\xspace}
\newcommand{\onethree}        {$\sqrt{s}~=~13$~Te\kern-.1emV\xspace}
\newcommand{\MeV}  {\ensuremath{\text{Me\kern-.1emV}}\xspace}
\newcommand{\MeVc}  {\ensuremath{\text{Me\kern-.1emV/}c}\xspace}
\newcommand{\MeVcc}  {\ensuremath{\text{Me\kern-.2emV/}c^2}\xspace}
\newcommand{\GeV}  {\ensuremath{\text{Ge\kern-.1emV}}\xspace}
\newcommand{\GeVc}  {\ensuremath{\text{Ge\kern-.1emV/}c}\xspace}
\newcommand{\GeVcc}  {\ensuremath{\text{Ge\kern-.2emV/}c^2}\xspace}
\newcommand{\TeV}  {\ensuremath{\text{Te\kern-.1emV}}\xspace}

\newcommand{\ITS}          {\rm{ITS}\xspace}
\newcommand{\TOF}          {\rm{TOF}\xspace}
\newcommand{\ZDC}          {\rm{ZDC}\xspace}
\newcommand{\ZDCs}         {\rm{ZDCs}\xspace}
\newcommand{\ZNA}          {\rm{ZNA}\xspace}
\newcommand{\ZNC}          {\rm{ZNC}\xspace}
\newcommand{\SPD}          {\rm{SPD}\xspace}
\newcommand{\SDD}          {\rm{SDD}\xspace}
\newcommand{\SSD}          {\rm{SSD}\xspace}
\newcommand{\TPC}          {\rm{TPC}\xspace}
\newcommand{\TRD}          {\rm{TRD}\xspace}
\newcommand{\TZERO}        {\rm{T0}\xspace}
\newcommand{\VZERO}        {\rm{V0}\xspace}
\newcommand{\VZEROA}       {\rm{V0A}\xspace}
\newcommand{\VZEROC}       {\rm{V0C}\xspace}
\newcommand{\Vdecay} 	   {\ensuremath{V^{0}}\xspace}

\newcommand{\phot}         {\ensuremath{\gamma}\xspace} 
\newcommand{\ee}           {\ensuremath{\mathrm{e}^{+}\mathrm{e}^{-}}\xspace} 
\newcommand{\pip}          {\ensuremath{\uppi^{+}}\xspace}
\newcommand{\pim}          {\ensuremath{\uppi^{-}}\xspace}
\newcommand{\kap}          {\ensuremath{\rm{K}^{+}}\xspace}
\newcommand{\kam}          {\ensuremath{\rm{K}^{-}}\xspace}
\newcommand{\pbar}         {\ensuremath{\rm\overline{p}}\xspace}
\newcommand{\kzero}        {\ensuremath{{\rm K}^{0}_{\rm{S}}}\xspace}
\newcommand{\lmb}          {\ensuremath{\upLambda}\xspace}
\newcommand{\sigm}         {\ensuremath{\upSigma}\xspace}
\newcommand{\almb}         {\ensuremath{\overline{\upLambda}}\xspace}
\newcommand{\Om}           {\ensuremath{\upOmega^-}\xspace}
\newcommand{\Mo}           {\ensuremath{\overline{\upOmega}^+}\xspace}
\newcommand{\X}            {\ensuremath{\upXi^-}\xspace}
\newcommand{\Ix}           {\ensuremath{\overline{\upXi}^+}\xspace}
\newcommand{\Xis}          {\ensuremath{\upXi^{\pm}}\xspace}
\newcommand{\Oms}          {\ensuremath{\upOmega^{\pm}}\xspace}
\newcommand{\degr}         {\ensuremath{^{\rm o}}\xspace}

\newcommand{\siZ}          {\ensuremath{\upSigma^{0}}\xspace}
\newcommand{\siP}          {\ensuremath{\upSigma^{+}}\xspace}
\newcommand{\siM}          {\ensuremath{\upSigma^{-}}\xspace}
\newcommand{\asiZ}         {\ensuremath{\overline{\upSigma^{0}}}\xspace}
\newcommand{\psiZCF}       {\rm{p}\mbox{--}\siZ}
\newcommand{\SNCF}         {\rm{N}\mbox{--}\ensuremath{\upSigma}\xspace}
\newcommand{\LNCF}         {\rm{N}\mbox{--}\ensuremath{\upLambda}\xspace}
\newcommand{\apasiZCF}     {\ensuremath{\overline{\mathrm{p}}}\mbox{--}\asiZ}
\newcommand{\psiZCombCF}   {\rm{p}\mbox{--}\siZ \ensuremath{\oplus} \apasiZCF}
\newcommand{\SB}           {\ensuremath{(\upLambda\gamma)}\xspace}
\newcommand{\pSBCF}        {\rm{p}\mbox{--}\SB}

\newcommand{\pLCF}         {\rm{p}\mbox{--}\lmb}
\newcommand{\nSplusCF}     {\rm{n}\mbox{--}\ensuremath{\upSigma^{+}}}

\newcommand{\Ledn}         {Lednick\'y--Lyuboshits approach\xspace}
\newcommand{\chiEFT}       {\ensuremath{\chi}\rm{EFT}\xspace}
\newcommand{\fss}       {\rm{fss2}\xspace}
\newcommand{\ESC}          {\rm{ESC16}\xspace}
\newcommand{\NSC}          {\rm{NSC97f}\xspace}

\newcommand{\pP}           {\ensuremath{\mbox{p--p}}\xspace}
\newcommand{\ApAP}         {\ensuremath{\mbox{\pbar--\pbar}}\xspace}
\newcommand{\pPComb}       {\ensuremath{\mbox{p--p} \oplus \mbox{\pbar--\pbar}}\xspace}

\newcommand{\radiusResult} {\ensuremath{r_0 = 1.249 \pm 0.008\, \mathrm{(stat)} \,^{+0.024} _{-0.021}\, \mathrm{(syst)}}\,fm\xspace}

\newcommand{\pppCF}         {\rm{p}\mbox{--}\rm{p}\mbox{--}\rm{p}\xspace}
\newcommand{\ppLCF}         {\rm{p}\mbox{--}\rm{p}\mbox{--}\lmb}
\newcommand{\apapapCF}         {\pbar\mbox{--}\pbar\mbox{--}\pbar\xspace}
\newcommand{\apapaLCF}         {\pbar\mbox{--}\pbar\mbox{--}\almb}

\newcommand{\ppspCF}         {(\rm{p}\mbox{--}\rm{p})\mbox{--}\rm{p}\xspace}
\newcommand{\ppsLCF}         {(\rm{p}\mbox{--}\rm{p})\mbox{--}\lmb\xspace}
\newcommand{\ppLsCF}         {\rm{p}\mbox{--}(\rm{p}\mbox{--}\lmb)\xspace}

\newcommand{\pppbarCF}         {\rm{p}\mbox{--}\rm{p}\mbox{--}\pbar\xspace}

\newcommand{\ppbarCF}         {\rm{p}\mbox{--}\pbar\xspace}

\newcommand{\ppCF}         {\rm{p}\mbox{--}\rm{p}\xspace}

\newcommand{\ppspbarCF}         {(\rm{p}\mbox{--}\rm{p})\mbox{--}\pbar\xspace}

\newcommand{\pppbarsCF}         {(\rm{p}\mbox{--}\pbar)\mbox{--}\rm{p}\xspace}

\newcommand{\Qthree}           {\ensuremath{Q_3}\xspace}


\newcommand{\ppKp}{\rm{p}\mbox{--}\rm{p}\mbox{--}\rm{K$^+$}\xspace}
\newcommand{\appKp}{\rm{\pbar}\mbox{--}\rm{\pbar}\mbox{--}\rm{K$^-$}\xspace}

\newcommand{\ppKm}{\rm{p}\mbox{--}\rm{p}\mbox{--}\rm{K$^-$}\xspace}
\newcommand{\appKm}{\rm{\pbar}\mbox{--}\rm{\pbar}\mbox{--}\rm{K$^+$}\xspace}

\newcommand{\ppsKp}{$\mathrm{(p-p)-K^{+}}$}
\newcommand{\appsKm}{$\mathrm{(\overline{p}-\overline{p})-K^{-}}$}

\newcommand{\ppKps}{$\mathrm{p-(p-K^{+})}$}
\newcommand{\appKms}{$\mathrm{\overline{p}-(\overline{p}-K^{-}})$ }

\newcommand{\ppsKm}{$\mathrm{(p-p)-K^{-}}$}
\newcommand{\appsKp}{$\mathrm{(\overline{p}-\overline{p})-K^{+}}$}

\newcommand{\ppKms}{$\mathrm{p-(p-K^{-})}$}
\newcommand{\appKps}{$\mathrm{\overline{p}-(\overline{p}-K^{+}})$ }

\newcommand{\ls}[1]{{\color{red}[LS: #1]}}

\newcommand{\geant}        {${\textsc{Geant4}}$\xspace}
\begin{titlepage}
\PHyear{2023}
\PHnumber{051}      
\PHdate{21 March}  

\title{Study of the p--p--K$^+$ and p--p--K$^-$ dynamics using the femtoscopy technique} \ShortTitle{p--p--K femtoscopy} 

\Collaboration{ALICE Collaboration\thanks{See Appendix~\ref{app:collab} for the list of collaboration members}}
\ShortAuthor{ALICE Collaboration} 

\begin{abstract}
The interactions of kaons (K) and antikaons ($\mathrm{\overline{K}}$) with few nucleons (N) were studied so far using kaonic atom data and measurements of kaon production and interaction yields in nuclei.  
Some details of the three-body KNN and $\mathrm{\overline{K}}$NN dynamics are still not well understood, mainly due to the overlap with multi-nucleon interactions in nuclei.
An alternative method to probe the dynamics of three-body systems with kaons is to study the final state interaction within triplet of particles emitted in pp collisions at the Large Hadron Collider, which are free from effects due to the presence of bound nucleons. 
This Letter reports the first femtoscopic study of \ppKp and \ppKm correlations measured in high-multiplicity pp collisions at $\sqrt{s}$ = 13 TeV by the ALICE Collaboration. The analysis shows that the measured \ppKp and \ppKm correlation functions can be interpreted in terms of pairwise interactions in the triplets, indicating that the dynamics of such systems is dominated by the two-body interactions without significant contributions from three-body effects or bound states.  
\end{abstract}
\end{titlepage}
\setcounter{page}{2}

The nature of kaons and antikaons is closely related to the chiral symmetry breaking pattern of low-energy QCD~\cite{Tolos:2020aln,Hyodo:2022xhp}. Therefore, the study of kaon properties and their modification in dense nuclear matter has received the attention of the scientific community in the past decades.
In particular, the hadronic interactions of charged kaons, K$^+$ and K$^-$, with nucleons (N) were investigated using kaonic atoms~\cite{BATTY1997385,friedman2017k,curceanu2019modern} and by studying the interaction or the production of kaons in light~\cite{PhysRevD.1.1267,BC,MICHAEL199629,PhysRevC.92.045204,doce2016k,Piscicchia:2018rez,del2019hbox,PhysRevC.51.669,PhysRevC.49.2569} and heavy nuclei~\cite{20006,PhysRevC.75.024906,PhysRevC.90.025210,PhysRevC.65.014603,PhysRevLett.96.072301,PhysRevLett.123.022002,ANKE,PhysRevC.99.014904,PhysRevLett.102.182501}. Such measurements demonstrated the repulsive and attractive nature of the K$^+$N and K$^-$N strong interactions~\cite{Tolos:2020aln,Hyodo:2022xhp}, respectively. Information on the in-medium modification of the K$^+$ and K$^-$ potentials with the increasing baryon density was extracted from the comparison of kaon production yields and flow observables measured in nucleus--nucleus collisions~\cite{20006,PhysRevC.75.024906,PhysRevC.90.025210} and pion- and proton-induced reactions~\cite{PhysRevC.65.014603,PhysRevLett.96.072301,PhysRevLett.123.022002,PhysRevLett.102.182501,ANKE,PhysRevC.99.014904} with the expectations from transport models. Several K$^+$ nuclear potentials have been tested and the best agreement with the data results in a repulsive strength of 20--40 MeV at nuclear saturation density. 
On the other hand, the K$^-$N interaction is known to be sufficiently attractive in the isospin I~=~0 channel to dynamically generate the $\Lambda(1405)$ just below the K$^-$p threshold~\cite{Mai2021}. Such a state is interpreted as a quasi-bound antikaon--nucleon $\mathrm{\overline{K}}$N system which couples strongly to the $\pi\Sigma$ channel, giving rise to a sizable K$^-$ absorption~\cite{SekiharaPRC}.
In the case of K$^-$ nuclear interaction, the influence of single- and multi-nucleon absorption processes in nuclei currently prevents extracting firm conclusions on the strength of the K$^-$ in-medium attractive potential~\cite{Tolos:2020aln}.
This ambiguity triggered a longstanding debate in the literature about the possible existence of exotic kaonic bound objects with nucleons (see~\cite{Tolos:2020aln} and references therein).
Recently, the E15 Collaboration reported the first experimental evidence of the $\mathrm{\overline{K}}$NN state with a binding energy of about 42 MeV and a decay width of about 100 MeV~\cite{yamaga2020observation}.
From the theoretical side, 
 binding energies in the range 9--95 MeV and decay widths between 16 and 110 MeV are expected~\cite{yamazaki2002k,shevchenko2007faddeev,shevchenko2007k,ikeda2007strange,dote2008k,ikeda2009resonance,wycech2009variational,koike2009k,ikeda2010energy,barnea2012realistic,bayar2013k,revai2014faddeev,dote2015application,sekihara2016structure,ohnishi2017few,dote2017fully}.
The uncertainties in the models mainly arise from scarce knowledge of the full $\mathrm{\overline{K}}$NN three-body effects, such as three-body coupled channels and two-nucleon absorption processes.
Additionally, three-body forces, which are relevant in the calculation of the nuclear binding energies~\cite{RevModPhys.85.197,marcucci2020hyperspherical}, are currently not included in kaonic bound state models. 
Further experimental investigations on the K$^+$NN and K$^-$NN three-body dynamics, in addition to the studies of K$^+$ and K$^-$ interaction in nuclei, are required to isolate and quantify the contribution from genuine three-body effects in such systems.

An alternative method to explore the three-body dynamics of K$^+$ and K$^-$ with nucleons is to employ the femtoscopy technique at high-energy collider facilities. In small colliding systems, such as pp and p--Pb at the Large Hadron Collider (LHC),
the inter-hadron distances at the time of the particle emission range
from a few femtometers down to scales compatible with the nucleon size. This leads to an enhancement of the strength of the signal due to the short-range strong interaction in the measured correlation function~\cite{Fabbietti:2020bfg}. The femtoscopy method was proven to be able to test and constrain the hadron--hadron interaction for various two-particle systems~\cite{ALICE:Run1,Acharya:2019bsa,acharya2020investigation,ALICE:LL, ALICE:pXi,ALICE:pOmega,ALICE:2021cpv,ALICE:2021cyj,ALICE:2022yyh,ALICE:2021njx,ALICE:2022uso,ALICE:2022enj}, providing data with unprecedented precision on the hadronic interactions with strangeness at low relative momenta, down to the energy threshold of the produced pairs. Furthermore, the measured femtoscopic correlation functions are sensitive to the presence of bound states in the energy region below the threshold. Thus, they were also used to constrain the parameters of bound hadron--hadron systems~\cite{ALICE:LL,Chizzali:2022pjd}. Recently, the ALICE Collaboration has extended the method to explore the correlation among three baryons, such as \pppCF and \ppLCF, to study the dynamics of three-body systems~\cite{ALICE:2022vzr}. The analysis exploited Kubo's cumulant expansion method~\cite{Kubo} and the projector method~\cite{DelGrande:2021mju} to isolate the genuine three-particle correlations from the measured correlation functions. As a result, a first experimental hint of genuine three-body effects in the unbound \pppCF system was found. The study in this Letter applies the same analysis procedure adopted in~\cite{ALICE:2022vzr} to the case of \ppKp and \ppKm particle triplets to explore possible genuine three-body effects in the correlation functions induced by the strong interaction and bound state formation. The main advantage of the femtoscopy method with respect to the previous experimental techniques is the possibility to investigate for the first time the K$^+$ and K$^-$ three-body dynamics with nucleons, free from additional effects induced by the presence of the surrounding nucleons in nuclei.

The analysed data sample of  \pp collisions at \thirteen was recorded by ALICE~\cite{ALICE, ALICEperf,ALICE:2022wpn} during the LHC Run 2 (2015--2018) data-taking period. In the following, the information from the V0 detector system~\cite{Abbas:2013taa}, the Inner Tracking System (ITS)~\cite{ALICEITS}, the Time Projection Chamber (TPC)~\cite{TPC} and the Time-Of-Flight (TOF) detector~\cite{TOF} is used. The V0 detector is employed to trigger on high-multiplicity (HM) events. This trigger selects events within the 0.17\% largest charged-track multiplicity of the INEL~$>0$ class, which is defined as inelastic collisions with at least one measured charged particle in the pseudorapidity range $|\eta|<1$~\cite{Abbas:2013taa}. This condition results in an average of about 30 charged particles in the range $|\eta|<0.5$~\cite{ALICE:pOmega}, hence increasing the probability of finding triplets of the desired particle species with respect to the minimum-bias sample.
The primary vertex position is reconstructed with the combined information of the ITS
and the TPC, and, independently, with track segments in the two innermost layers of the ITS.
Only events with a reconstructed primary vertex position along the beam axis within
10 cm from the centre of the ALICE detector are selected.
A total of 10$^9$ HM events are used in this analysis. 
The \ppKp and \ppKm data samples are built 
by combining three charged-particle tracks (triplets) reconstructed with the TPC.  
Assuming the same interactions in the particle and antiparticle systems, triplets of particles and the corresponding antiparticles are combined: \ppKp $\equiv$ (~\ppKp~$\oplus$~\appKp) and \ppKm $\equiv$ (~\ppKm~$\oplus$~\appKm).
The agreement of the corresponding correlation functions confirmed the validity of this assumption.

Particles and antiparticles are identified using the same kinematic and topological selections.  
Protons and kaons are selected in the range $|\eta| < 0.8$ and in the transverse momentum intervals $\pt \in$ (0.5--4.05)\,\GeVc and $\pt \in$ (0.2--2.5)\,\GeVc, respectively. 
A minimum of 80 space points (hits) inside the TPC, out of the total 159, is required to guarantee track quality and good momentum resolution. 
Particle identification (PID) is conducted by requiring that the measured energy loss (\dEdx) in the TPC gas is compatible with the expected one from protons and kaons within three standard deviations ($\sigma$).         
For high momentum particles, the \dEdx information is combined with the time-of-flight measurement provided by the TOF, using a 3$\sigma$ selection on the expected value for a given particle hypothesis at a given momentum.
The PID selection for protons is described in detail in~\cite{ALICE:Run1}. The selection of kaons is based on the procedure described in ~\cite{ALICE:2022yyh} with several changes as explained in the following. 
TPC reconstructed tracks are identified as kaons either by using TPC PID information for momenta lower than 0.85 GeV/$c$, or, if a signal in the TOF is matched to the TPC track, by using the combined PID information from TPC and TOF up to a momentum of 2.5 GeV/$c$. 
To improve the purity of the kaon selection, the TPC and TOF information are also used to reject candidates that are compatible with the pion or electron hypothesis.
To reject particles that are non-primary or come from pile-up collisions, the distance of closest approach (DCA) to the primary vertex of the tracks is required to be less than 0.1~cm in the transverse plane and less than 0.2~cm along the beam axis. The purity of the proton and kaon candidates is estimated employing Monte Carlo simulations based on PYTHIA 8~\cite{Sjostrand:2014zea} (Monash 2013 Tune) event generator with a dedicated high-multiplicity selection to mimic the V0 high-multiplicity trigger and the \geant package~\cite{Agostinelli:2002hh,Uzhinsky:2011zz}. The purity averaged over the \pt ranges of the identified protons and kaons is about $98\%$ for both particle species.

The number of selected and analysed triplets amounts to 4530 for \ppKp, 3161 for \appKp, 6200 for \ppKm and 4937 for \appKm in the femtoscopic region $Q_3 < 0.4$ \GeVc. The kinematic variable $Q_3$, which is used in three-body analyses~\cite{ALICE:2022vzr,Dhevan}, is a Lorentz-invariant scalar defined as $Q_3=\sqrt{-q_{12}^2-q_{23}^2 -q_{31}^2}$, where $q_{ij}$ is the norm of the relative four-momentum of the pair $ij$ in the triplet $q_{ij}^\mu = 2\ [ m_j/(m_i~+~m_j) ~p_i^\mu~-~m_i/(m_i + m_j)\ p_j^\mu ]$. 
The systematic uncertainties of the correlation functions are evaluated by performing simultaneous variations of the selection criteria for particles and antiparticles. For protons and antiprotons, the same variations for the track selection and PID criteria as in~\cite{ALICE:2022vzr} are performed. Similar variations are used for the kaons and antikaons. In order to account for the correlations between the systematic uncertainties, the variations are randomly combined in sets in which at least one selection criterion is changed. Large statistical fluctuations introduced by the variations of the selection criteria are avoided by requiring that 
the yield of the triplets changes by less than 10\% with respect to the standard selection in the femtoscopic region $Q_3 < 0.4$ \GeVc. The main contribution at low $Q_3$ ($\approx 0.15$ \GeVc) is given by the variation of the DCA selection and it is found to be smaller than 5\%.

The final state interactions (FSIs) among the hadrons emitted in the collisions can be explored using correlation functions~\cite{lednicky2004correlation,Lisa:2005dd} in momentum space, defined 
as
\begin{equation}
   C_3(\mathbf{p}_1,\mathbf{p}_2,\mathbf{p}_3) = \frac{P_3(\mathbf{p}_1,\mathbf{p}_2,\mathbf{p}_3)}{P_1(\mathbf{p}_1) P_1(\mathbf{p}_2) P_1(\mathbf{p}_3)} \ ,
\end{equation}
where $\mathbf{p}_i$ is the momentum vector of the $i$-th particle and $P_3(\mathbf{p}_1,\mathbf{p}_2,\mathbf{p}_3)$ and $P_1(\mathbf{p}_i)$  are the probabilities of finding three particles and one particle with the corresponding momentum, respectively.
If the particles are not correlated $P_3(\mathbf{p}_1,\mathbf{p}_2,\mathbf{p}_3) = P_1(\mathbf{p}_1) P_1(\mathbf{p}_2) P_1(\mathbf{p}_3)$ and then the correlation function becomes identical to unity.
The presence of FSIs induces a correlation signal imprinted in the correlation function which causes deviations from unity depending on the nature of the interactions, attractive or repulsive, as well as on the properties of the emitting source (see~\cite{Fabbietti:2020bfg,ALICE:2020ibs} for the details).
In the case of three-particle systems, the correlation function can be evaluated in terms of $Q_3$. Hence, three-particle correlation functions are experimentally obtained as 
\begin{equation}
    C_3(Q_3) = \mathcal{N}\ \frac{N_\mathrm{same}(Q_3)}{N_\mathrm{mixed}(Q_3)} \ ,
\end{equation}
where $\mathcal{N}$ is a normalisation parameter, $N_\mathrm{same}(Q_3)$ is the $Q_3$ distribution of the particle triplets emitted in the same collision and  $N_\mathrm{mixed}(Q_3)$ is the distribution of uncorrelated triplets. The latter are obtained by taking the three particles needed to form a triplet from three different collisions, thus called mixed event sample.
Only events with similar multiplicity and position of the primary vertex along the beam axis are mixed.
The number of events used for the mixing is set to 30. To account for the two-track merging and splitting effects due to the finite two-track resolution in the same event sample, a minimum value of the distance between the same charge tracks on the azimuthal--polar angle plane $\Delta \eta$--$\Delta \varphi$ is applied to both the same and mixed event samples. The default selection is $\Delta\eta^2 + \Delta \varphi^2 \geq (0.017)^2$ for~ p--p tracks and $\Delta\eta^2/(0.012)^2 + \Delta \varphi^2/(0.04)^2 \geq 1$ for p--K$^+$ tracks. A systematic variation of about $+$10\% for the values of the minimum distance is applied in the analysis.
The correlation function is normalised to unity in the range $1.0 < Q_3 < 1.2$ GeV/$c$, where no signal from FSIs is expected.
Variations of the normalisation range are performed and included in the systematic uncertainties. 
In the case of a particle triplet X--Y--Z, pairwise correlations with one uncorrelated particle (X--Y)--Z, X--(Y--Z) and (Z--X)--Y also occur in the system. Genuine three-particle correlations were isolated in~\cite{ALICE:2022vzr} by applying the cumulant expansion method to the femtoscopic correlation functions. The three-particle cumulant $c_3 (Q_3)$, which incorporates information about the three-body effects in the system, is computed from the measured correlation function $C_3(Q_3)$ as follows
\begin{equation}
    c_3 (Q_3) = C_{3}(Q_3) - C_\mathrm{two\text{-}body}(Q_3) \ ,
    \label{eq:cumulant}
\end{equation}
where pairwise correlations are evaluated following the cumulant decomposition as
\begin{equation}
    C_\mathrm{two\text{-}body}(Q_3) = C_\mathrm{{}_{(X-Y)-Z}}(Q_3) + C_\mathrm{{}_{X-(Y-Z)}}(Q_3) + C_\mathrm{{}_{(Z-X)-Y}}(Q_3) - 2  \ .
    \label{eq:loc}
\end{equation}
Each component $C_\mathrm{{}_{(i-j)-k}}(Q_3)$ of the lower-order contributions in Eq. \eqref{eq:loc} is computed using two methods~\cite{ALICE:2022vzr,DelGrande:2021mju}: $i)$ a data-driven approach based on the event mixing technique, building triplets in which the correlated (i--j) pairs are emitted in the same collision and the uncorrelated particle --k from another collision; $ii)$ the projector method which uses as input the measured two-particle correlation function $C(k^*)$ for the correlated pair (i--j), evaluated in terms of the relative momentum $k^*$ in the pair rest frame (PRF), and calculates all the possible $k^*$ configurations in the phase space for each $Q_3$ value of the i--j--k triplet. The input correlation functions $C(k^*)$ used in the projector method are obtained by selecting p--p,  p--K$^+$ and p--K$^-$ pairs using the same dataset as for the three-body analysis. Unlike the p--p and p--K$^+$ pairs, the p--K$^-$ correlation is significantly affected by the contribution from jet-like events~\cite{ALICE:2011kmy}. To use particles emitted in collisions with the same event shape as in the case of the analysed triplets, p--K$^-$ pairs are selected from \ppKm triplets with $Q_3 < 1$ GeV/$c$, while for p--p and p--K$^+$ pairs such additional requirement is not used as the correlation functions with and without the $Q_3$ selection are in agreement. A $\pm 0.1$ GeV/$c$ variation of the $Q_3$ limit is included in the systematic uncertainties of the projector method and it is combined with the uncertainties propagated from the data points of the p--p, p--K$^+$ and p--K$^-$ correlation functions. The latter include the variations in the selection criteria for particles, variations in the track splitting/merging rejection criteria, and normalisation range.

\begin{figure}[!h]
    \centering
        \includegraphics[width=0.48\textwidth]{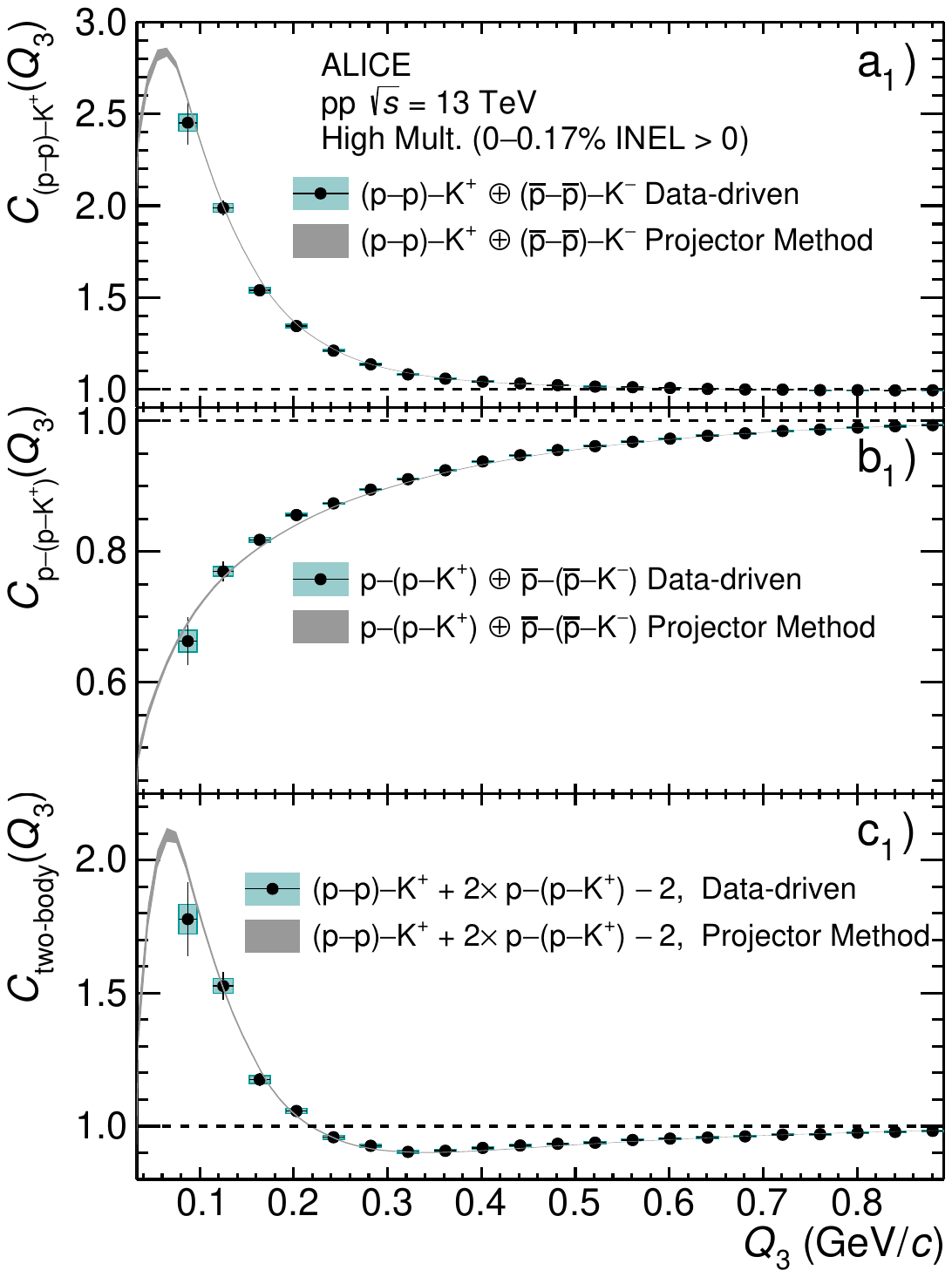}
    \includegraphics[width=0.48\textwidth]{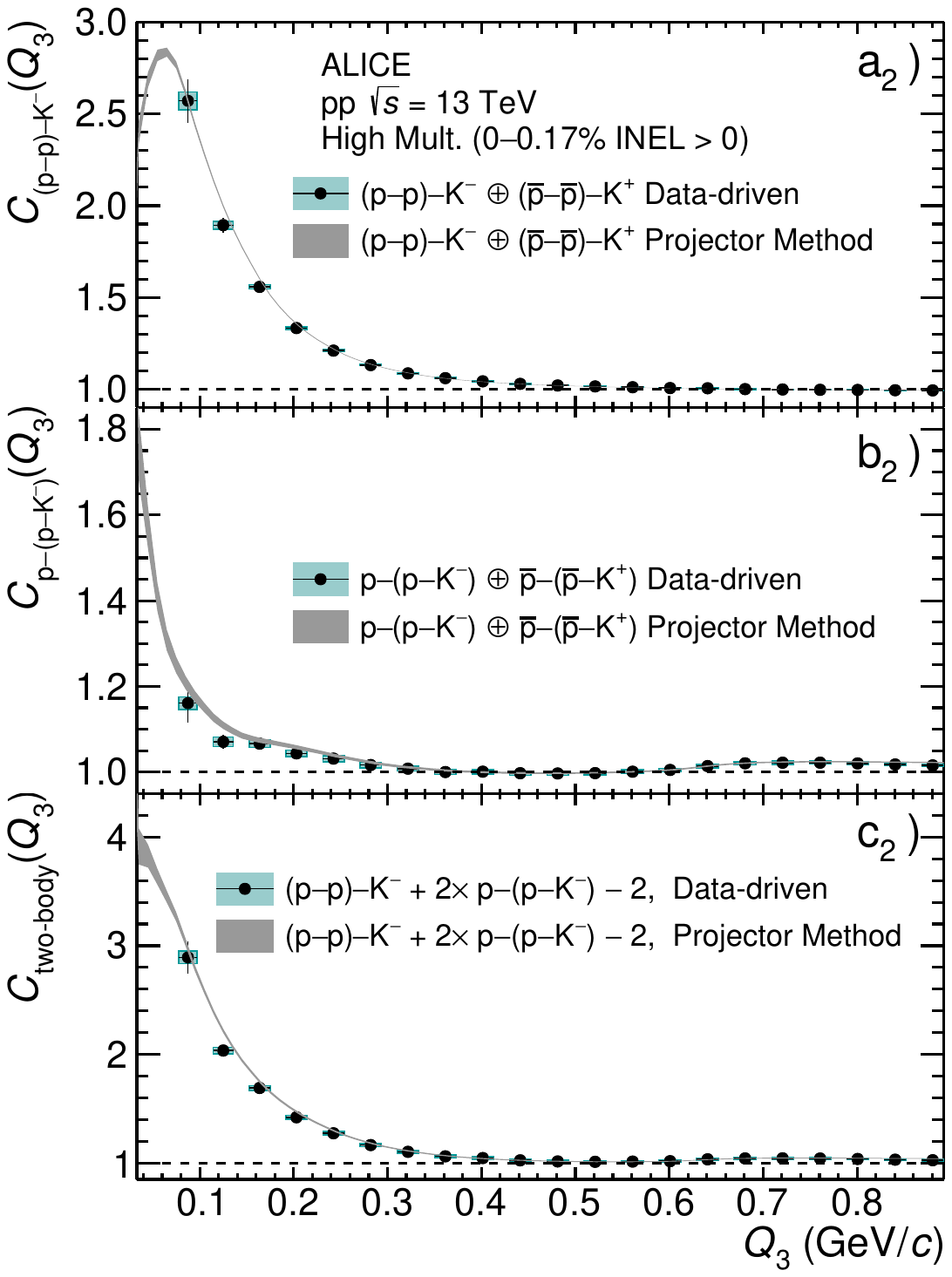}
    \caption{Correlation functions for (p--p)--K$^+$ (panel $\mathrm{a_1}$), (p--p)--K$^-$ (panel $\mathrm{a_2}$), p--(p--K$^+$) (panel $\mathrm{b_1}$), p--(p--K$^-$) (panel $\mathrm{b_2}$) as well as the total lower-order contributions for \ppKp (panel $\mathrm{c_1}$) and \ppKm (panel $\mathrm{c_2}$) as a function of $Q_3$. The points represent the results obtained using the data-driven approach, with the statistical and systematic uncertainties represented by the error bars and the green boxes, respectively. The grey bands are the expectations of the projector method, with the band width representing the combined statistical and systematic uncertainties.}
    \label{fig:locPPK}
\end{figure}
\begin{figure}[!h]
    \centering
    \includegraphics[width=0.495\textwidth]{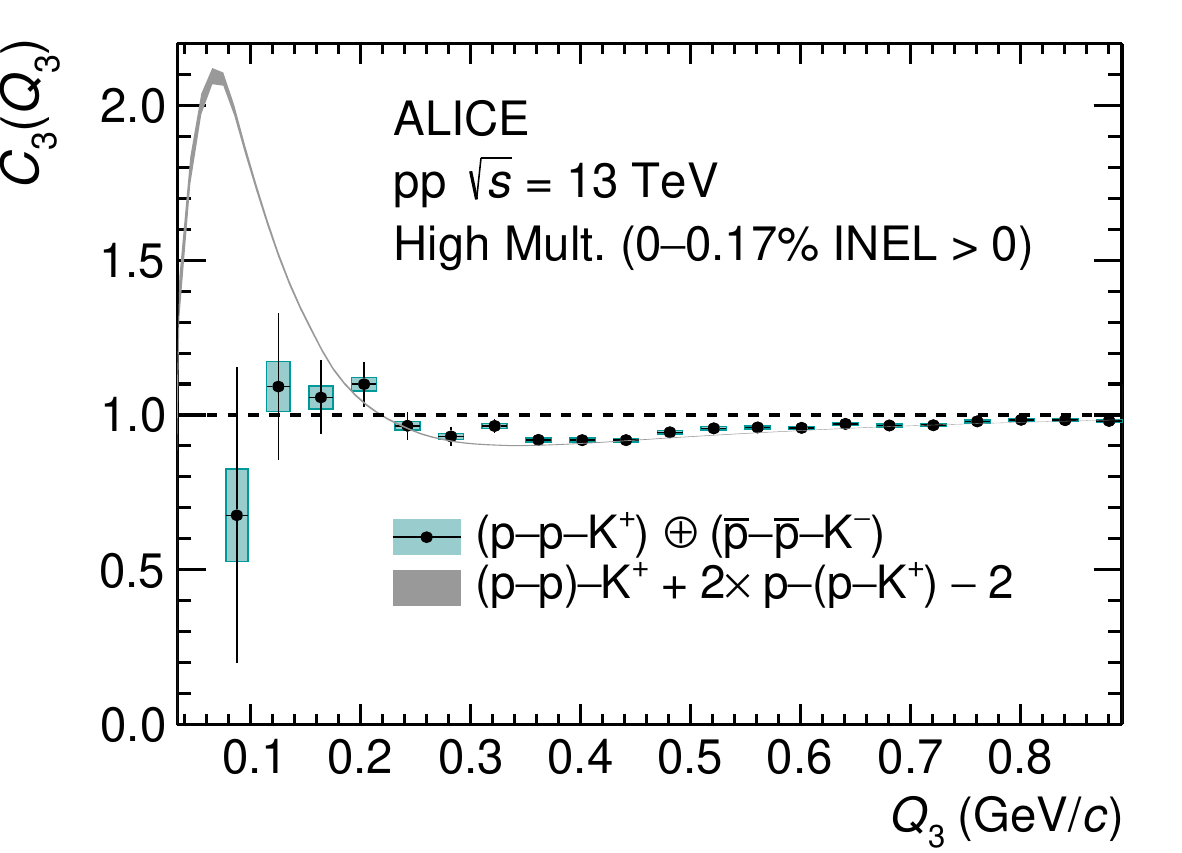}
    \includegraphics[width=0.495\textwidth]{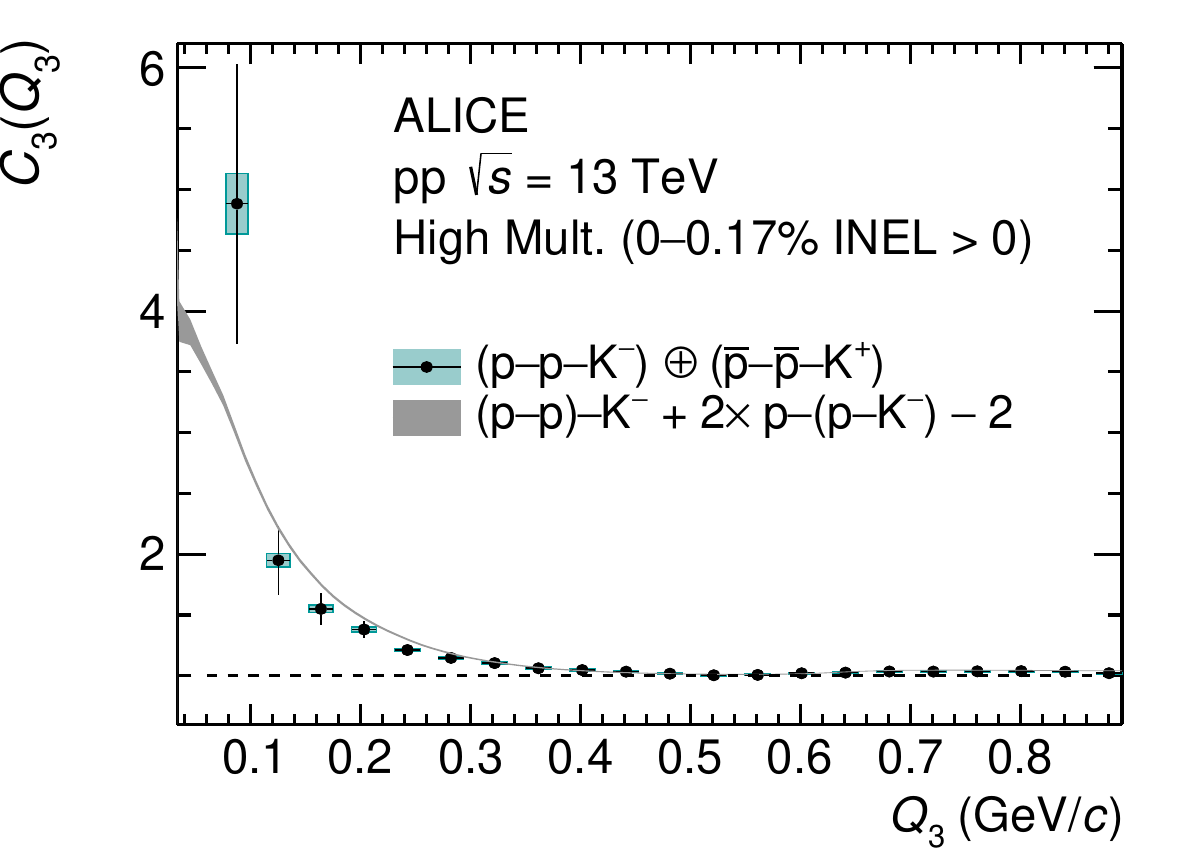}
    \caption{Correlation functions (data points) for \ppKp (left panel) and \ppKm (right panel) compared to the lower-order contributions evaluated using the projector method (grey bands). Statistical and systematic uncertainties are represented by error bars and green boxes, respectively. The band widths represent the combined statistical and systematic uncertainties propagated from the two-particle correlation functions used as input to the projector method. }
    \label{fig:CFs}
\end{figure}

For the \ppKp and \ppKm systems, the lower-order contributions to the three-particle correlation functions are shown in Fig.~\ref{fig:locPPK}, using the data-driven approach (data points) and the projector method (grey bands). 
The three-particle correlation functions of two correlated protons and an uncorrelated kaon, denoted as (p--p)--K$^+$ and (p--p)--K$^-$, are shown in panels $\mathrm{a_1}$) and $\mathrm{a_2}$), respectively. 
In both cases, the correlation functions are larger than unity in the low $Q_3$ region as a result of the attractive p--p strong interaction, combined with the repulsive Coulomb and quantum statistics effects (see~\cite{ALICE:Run1,acharya2020investigation} for more details).
In the case of correlated p--K$^+$ pairs with uncorrelated protons, p--(p--K$^+$), the correlation function shown in panel $\mathrm{b_1}$) is lower than unity as a result of the repulsive p--K$^+$ strong and Coulomb interactions, consistent with the measurement reported in~\cite{Acharya:2019bsa}.
In the p--(p--K$^-$) correlation function, shown in panel $\mathrm{b_2}$), the main features of the p--K$^-$ interaction are visible~\cite{Acharya:2019bsa}: the cusp structure due to the opening of the $\mathrm{\overline{K}^0}$n channel as well as the bump due to the $\mathrm{\Lambda(1520) \rightarrow p K^{-}}$ decay appear at $Q_3 \approx$ 0.15 GeV/$c$ and $Q_3 \approx$ 0.7 GeV/$c$, respectively.
The total lower-order contributions to the \ppKp, panel $\mathrm{c_1}$), and \ppKm, panel $\mathrm{c_2}$), correlation functions 
evaluated with the data-driven approach and the projector method are in agreement. 
The reduced $\chi^2$ between the two methods, evaluated in the $Q_3$ range shown in Fig.~\ref{fig:locPPK} (21 degrees of freedom) by adding in quadrature the statistical and systematic uncertainties of the data, is 1.6 and 1.4 for \ppKp and \ppKm, respectively. Since the projector method does not depend on the third particle mixing, it provides significantly smaller uncertainties than the data-driven approach and hence is used to extract the cumulants. This choice does not affect the final results of the analysis.
Figure~\ref{fig:CFs} shows the measured \ppKp (left panel) and \ppKm (right panel) correlation functions (data points) compared to the lower-order contributions evaluated using the projector method (grey bands). 

\begin{figure}
    \centering
    \includegraphics[width=0.495\textwidth]{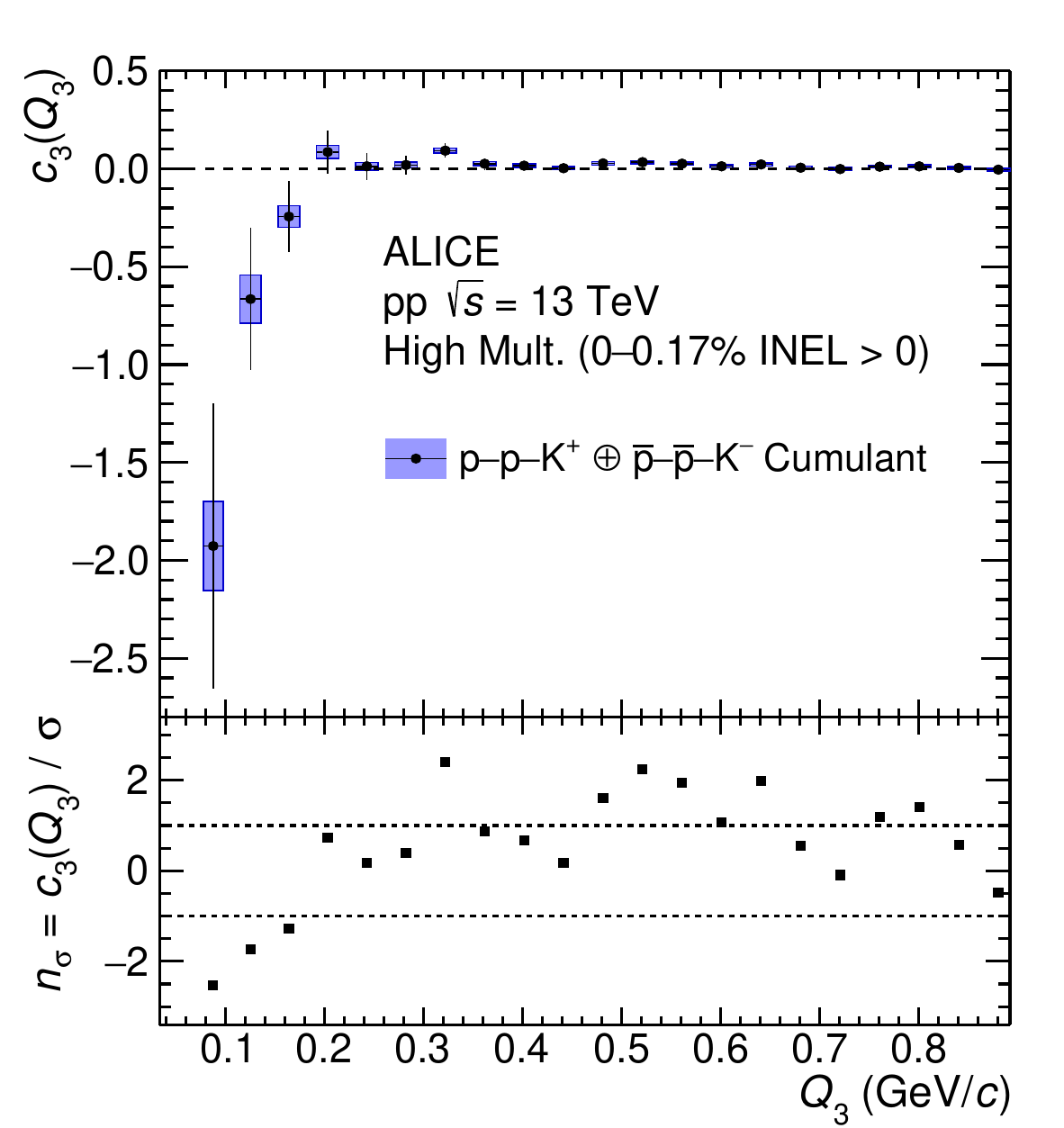}
    \includegraphics[width=0.495\textwidth]{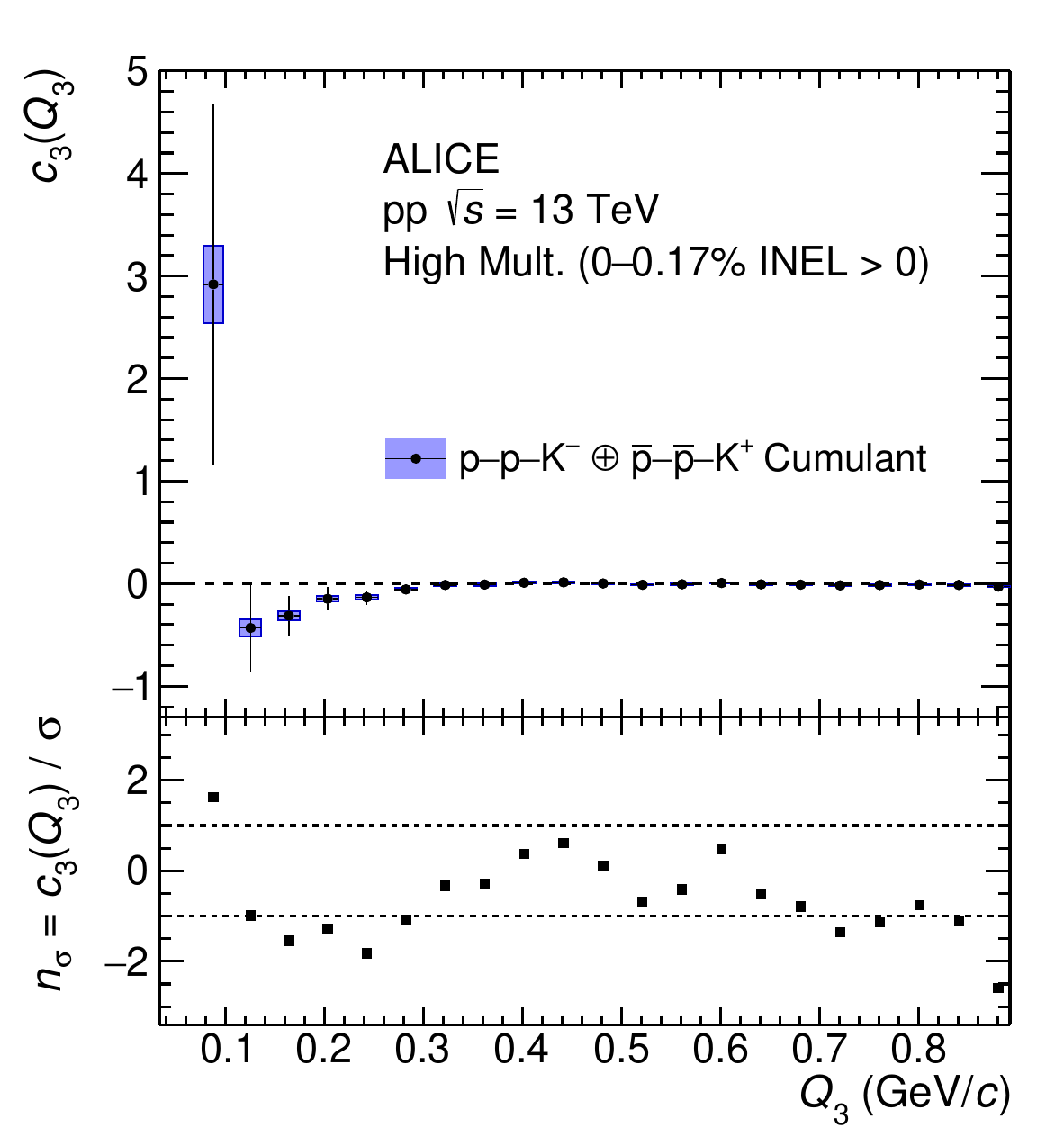}
    \caption{Cumulants for the \ppKp (left panel) and \ppKm (right panel) primary triplets. The error bars and the blue boxes on the data points represent the statistical and systematic uncertainties, respectively. The $n_\sigma$ deviations from zero in each bin are shown in the bottom panels, adding in quadrature the statistical and systematic uncertainties of the data.}
    \label{fig:cumulants}
\end{figure}

The corresponding cumulants are extracted using Eq. \eqref{eq:cumulant}. 
The measured cumulants include the correlations of the primary particles as well as the feed-down from resonances and particle misidentifications, which need to be accounted for as discussed in~\cite{ALICE:2022vzr}. 
The number of correctly identified primary triplets is about 66\% of the total sample for both \ppKp and \ppKm. The remaining contribution mainly stems from the feed-down of the $\Lambda$ (17\%) and $\Sigma^+$ (7\%) hyperon decays into protons and from the $\phi$ meson (4\%) decay into charged kaons. These numbers are obtained combining the primary and secondary fractions with particle purities. The primary and secondary fractions are extracted using Monte Carlo template fits to the measured distributions of the DCA to the primary vertex~\cite{ALICE:Run1}. The fraction of charged kaons from $\phi$ decays is calculated using the expectations from a thermal model~\cite{VOVCHENKO2019295}.  The correction of the measured cumulants to obtain the contribution from primary triplets is performed following the decomposition procedure adopted in~\cite{ALICE:2022vzr}. 
Due to the absence of theoretical and experimental information on the genuine p--$\Lambda$--K$^\pm$, p--$\Sigma^{+}$--K$^\pm$ and p--p--$\phi$ correlations, the corresponding feed-down contributions to the p--p--K$^{\pm}$ cumulants are considered to be flat, without any specific dependence on $Q_3$. This assumption was tested in the case of p--p--p correlations~\cite{ALICE:2022vzr}, where a similar contribution (about 19\%) for the $\Lambda$ feed-down into protons was found. The results obtained by using the flat and non-flat corrections are found to be in agreement within the uncertainties.
A flat feed-down contribution in the cumulant corresponds to negligible correlations from genuine three-body effects in the mother particle systems X--Y--Z decaying into p--p--K$^{\pm}$. The dominant feed-down contributions coming from pairwise correlations in the triplets, such as (X--Y)--Z and X--(Y--Z), are already removed as they are included in the lower-order correlations. 
The resulting cumulants of the correctly identified primary \ppKp and \ppKm particle triplets are shown in left and right panels of Fig.~\ref{fig:cumulants}, respectively.

In the absence of genuine three-particle correlations in the triplets, the measured three-particle correlation function would be identical to the lower-order contributions $C_{3}(Q_3) = C_\mathrm{two\text{-}body}(Q_3)$. Consequently, from Eq. \eqref{eq:cumulant}, the cumulant would be compatible with zero within the experimental uncertainties. In the case of the \ppKp and \ppKm systems, the agreement of the measured cumulants with zero is evaluated by performing a $\chi^2$ test in the region $Q_3 < 0.4$ GeV/$c$. In this region, the relative momentum for all the pairs in the particle triplets is lower than 0.2 GeV/$c$, which is the kinematic region sensitive to the strong interaction. Larger $Q_3$ values are dominated by kinematic configurations in which one of the pairs has a relative momentum larger than 0.2 GeV/$c$, meaning that only two of the three particles in the triplet can be in a kinematic configuration which is favourable for the strong interaction. Thus, at $Q_3 > 0.4$ GeV/$c$ the contributions from the two-body interaction dominate. The local statistical significance at $Q_3 < 0.4$ GeV/$c$ is obtained from the p-value of the $\chi^2$ distribution, which is converted in a number $n_\sigma$ of Gaussian standard deviations. As the systematic uncertainties in different $Q_3$ bins are correlated, the $\chi^2$ is calculated using the cumulants extracted from each systematic variation with the corresponding statistical uncertainties. Finally, the average $\chi^2$ in the interval $Q_3 < 0.4$ GeV/$c$ is extracted.
The corresponding $n_\sigma$ values are 2.5 and 1.5 for \ppKp and \ppKm, respectively. As the $Q_3$ range of the genuine three-body effects is not known a priori, the $n_\sigma$ values have been evaluated as well in the region $Q_3 < 0.2$ GeV/$c$ and they are found to be 2.7 and 1.4 for \ppKp and \ppKm, respectively.  Such results indicate that the measured \ppKp and \ppKm correlation functions are compatible, within the quoted significance levels, with the assumption of pairwise correlations in the triplets without additional contributions from genuine three-body effects. More solid conclusions require a larger data sample to reduce the statistical uncertainties and full-fledged theoretical calculations for the three-particle correlation functions.
In order to provide a comparison of the kinematic range accessible with this measurement to the study of K$^+$ and K$^-$ interactions in light nuclei, the momenta of the three particles in the triplets with $Q_3 < 0.4$ GeV/$c$ are evaluated in the PRF of the two protons. The momentum of each proton in the PRF is $p_\mathrm{p}^* < 180$ MeV/$c$, compatible with the typical Fermi momenta of nucleons in light nuclei (e.g.~the Fermi momentum in Carbon-12 is about 220 MeV/$c$). In the rest frame of the two protons, the momentum of the K$^+$ and K$^-$ ranges from 30 MeV/$c$ (in the first bin in Fig.~\ref{fig:cumulants}) to 130 MeV/$c$ (eighth bin in Fig.~\ref{fig:cumulants}). 
This demonstrates that even for low kaon momenta, i.e. at energies close to the \ppKp and \ppKm thresholds, three-body effects such as kaonic bound state formation below threshold or three-body interactions do not contribute significantly to the measured correlation functions. 
These results provide additional experimental information for theoretical models aiming to understand the role of genuine three-body effects in \ppKp and \ppKm systems. 
The LHC Run 3 data taking will deliver a larger data set for more detailed studies of the K$^+$ and K$^-$ three-body dynamics in the low momentum region, down to the energy threshold of the triplets.


\newenvironment{acknowledgement}{\relax}{\relax}
\begin{acknowledgement}
\section*{Acknowledgements}

The ALICE Collaboration would like to thank all its engineers and technicians for their invaluable contributions to the construction of the experiment and the CERN accelerator teams for the outstanding performance of the LHC complex.
The ALICE Collaboration gratefully acknowledges the resources and support provided by all Grid centres and the Worldwide LHC Computing Grid (WLCG) collaboration.
The ALICE Collaboration acknowledges the following funding agencies for their support in building and running the ALICE detector:
A. I. Alikhanyan National Science Laboratory (Yerevan Physics Institute) Foundation (ANSL), State Committee of Science and World Federation of Scientists (WFS), Armenia;
Austrian Academy of Sciences, Austrian Science Fund (FWF): [M 2467-N36] and Nationalstiftung f\"{u}r Forschung, Technologie und Entwicklung, Austria;
Ministry of Communications and High Technologies, National Nuclear Research Center, Azerbaijan;
Conselho Nacional de Desenvolvimento Cient\'{\i}fico e Tecnol\'{o}gico (CNPq), Financiadora de Estudos e Projetos (Finep), Funda\c{c}\~{a}o de Amparo \`{a} Pesquisa do Estado de S\~{a}o Paulo (FAPESP) and Universidade Federal do Rio Grande do Sul (UFRGS), Brazil;
Bulgarian Ministry of Education and Science, within the National Roadmap for Research Infrastructures 2020-2027 (object CERN), Bulgaria;
Ministry of Education of China (MOEC) , Ministry of Science \& Technology of China (MSTC) and National Natural Science Foundation of China (NSFC), China;
Ministry of Science and Education and Croatian Science Foundation, Croatia;
Centro de Aplicaciones Tecnol\'{o}gicas y Desarrollo Nuclear (CEADEN), Cubaenerg\'{\i}a, Cuba;
Ministry of Education, Youth and Sports of the Czech Republic, Czech Republic;
The Danish Council for Independent Research | Natural Sciences, the VILLUM FONDEN and Danish National Research Foundation (DNRF), Denmark;
Helsinki Institute of Physics (HIP), Finland;
Commissariat \`{a} l'Energie Atomique (CEA) and Institut National de Physique Nucl\'{e}aire et de Physique des Particules (IN2P3) and Centre National de la Recherche Scientifique (CNRS), France;
Bundesministerium f\"{u}r Bildung und Forschung (BMBF) and GSI Helmholtzzentrum f\"{u}r Schwerionenforschung GmbH, Germany;
General Secretariat for Research and Technology, Ministry of Education, Research and Religions, Greece;
National Research, Development and Innovation Office, Hungary;
Department of Atomic Energy Government of India (DAE), Department of Science and Technology, Government of India (DST), University Grants Commission, Government of India (UGC) and Council of Scientific and Industrial Research (CSIR), India;
National Research and Innovation Agency - BRIN, Indonesia;
Istituto Nazionale di Fisica Nucleare (INFN), Italy;
Japanese Ministry of Education, Culture, Sports, Science and Technology (MEXT) and Japan Society for the Promotion of Science (JSPS) KAKENHI, Japan;
Consejo Nacional de Ciencia (CONACYT) y Tecnolog\'{i}a, through Fondo de Cooperaci\'{o}n Internacional en Ciencia y Tecnolog\'{i}a (FONCICYT) and Direcci\'{o}n General de Asuntos del Personal Academico (DGAPA), Mexico;
Nederlandse Organisatie voor Wetenschappelijk Onderzoek (NWO), Netherlands;
The Research Council of Norway, Norway;
Commission on Science and Technology for Sustainable Development in the South (COMSATS), Pakistan;
Pontificia Universidad Cat\'{o}lica del Per\'{u}, Peru;
Ministry of Education and Science, National Science Centre and WUT ID-UB, Poland;
Korea Institute of Science and Technology Information and National Research Foundation of Korea (NRF), Republic of Korea;
Ministry of Education and Scientific Research, Institute of Atomic Physics, Ministry of Research and Innovation and Institute of Atomic Physics and University Politehnica of Bucharest, Romania;
Ministry of Education, Science, Research and Sport of the Slovak Republic, Slovakia;
National Research Foundation of South Africa, South Africa;
Swedish Research Council (VR) and Knut \& Alice Wallenberg Foundation (KAW), Sweden;
European Organization for Nuclear Research, Switzerland;
Suranaree University of Technology (SUT), National Science and Technology Development Agency (NSTDA), Thailand Science Research and Innovation (TSRI) and National Science, Research and Innovation Fund (NSRF), Thailand;
Turkish Energy, Nuclear and Mineral Research Agency (TENMAK), Turkey;
National Academy of  Sciences of Ukraine, Ukraine;
Science and Technology Facilities Council (STFC), United Kingdom;
National Science Foundation of the United States of America (NSF) and United States Department of Energy, Office of Nuclear Physics (DOE NP), United States of America.
In addition, individual groups or members have received support from:
European Research Council, Strong 2020 - Horizon 2020 (grant nos. 950692, 824093), European Union;
Academy of Finland (Center of Excellence in Quark Matter) (grant nos. 346327, 346328), Finland;
Programa de Apoyos para la Superaci\'{o}n del Personal Acad\'{e}mico, UNAM, Mexico.

\end{acknowledgement}

\bibliographystyle{utphys} 
\bibliography{main}

\newpage
\appendix

\section{The ALICE Collaboration}
\label{app:collab}
\begin{flushleft} 
\small

S.~Acharya\,\orcidlink{0000-0002-9213-5329}\,$^{\rm 125}$, 
D.~Adamov\'{a}\,\orcidlink{0000-0002-0504-7428}\,$^{\rm 86}$, 
A.~Adler$^{\rm 69}$, 
G.~Aglieri Rinella\,\orcidlink{0000-0002-9611-3696}\,$^{\rm 32}$, 
M.~Agnello\,\orcidlink{0000-0002-0760-5075}\,$^{\rm 29}$, 
N.~Agrawal\,\orcidlink{0000-0003-0348-9836}\,$^{\rm 50}$, 
Z.~Ahammed\,\orcidlink{0000-0001-5241-7412}\,$^{\rm 132}$, 
S.~Ahmad\,\orcidlink{0000-0003-0497-5705}\,$^{\rm 15}$, 
S.U.~Ahn\,\orcidlink{0000-0001-8847-489X}\,$^{\rm 70}$, 
I.~Ahuja\,\orcidlink{0000-0002-4417-1392}\,$^{\rm 37}$
% ...

\section*{Affiliation Notes}

$^{\rm I}$ Deceased\\
$^{\rm II}$ Also at: Max-Planck-Institut f\"{u}r Physik, Munich, Germany\\
$^{\rm III}$ Also at: Italian National Agency for New Technologies, Energy and Sustainable Economic Development (ENEA), Bologna, Italy\\
$^{\rm IV}$ Also at: Department of Applied Physics, Aligarh Muslim University, Aligarh, India\\
$^{\rm V}$ Also at: Institute of Theoretical Physics, University of Wroclaw, Poland\\
$^{\rm VI}$ Also at: An institution covered by a cooperation agreement with CERN\\

\section*{Collaboration Institutes}

$^{1}$ A.I. Alikhanyan National Science Laboratory (Yerevan Physics Institute) Foundation, Yerevan, Armenia\\
$^{2}$ AGH University of Science and Technology, Cracow, Poland\\
$^{3}$ Bogolyubov Institute for Theoretical Physics, National Academy of Sciences of Ukraine, Kiev, Ukraine\\
$^{4}$ Bose Institute, Department of Physics  and Centre for Astroparticle Physics and Space Science (CAPSS), Kolkata, India\\
$^{5}$ California Polytechnic State University, San Luis Obispo, California, United States\\
$^{6}$ Central China Normal University, Wuhan, China\\
$^{7}$ Centro de Aplicaciones Tecnol\'{o}gicas y Desarrollo Nuclear (CEADEN), Havana, Cuba\\
$^{8}$ Centro de Investigaci\'{o}n y de Estudios Avanzados (CINVESTAV), Mexico City and M\'{e}rida, Mexico\\
$^{9}$ Chicago State University, Chicago, Illinois, United States\\
$^{10}$ China Institute of Atomic Energy, Beijing, China\\
$^{11}$ Chungbuk National University, Cheongju, Republic of Korea\\
$^{12}$ Comenius University Bratislava, Faculty of Mathematics, Physics and Informatics, Bratislava, Slovak Republic\\
$^{13}$ COMSATS University Islamabad, Islamabad, Pakistan\\
$^{14}$ Creighton University, Omaha, Nebraska, United States\\
$^{15}$ Department of Physics, Aligarh Muslim University, Aligarh, India\\
$^{16}$ Department of Physics, Pusan National University, Pusan, Republic of Korea\\
$^{17}$ Department of Physics, Sejong University, Seoul, Republic of Korea\\
$^{18}$ Department of Physics, University of California, Berkeley, California, United States\\
$^{19}$ Department of Physics, University of Oslo, Oslo, Norway\\
$^{20}$ Department of Physics and Technology, University of Bergen, Bergen, Norway\\
$^{21}$ Dipartimento di Fisica, Universit\`{a} di Pavia, Pavia, Italy\\
$^{22}$ Dipartimento di Fisica dell'Universit\`{a} and Sezione INFN, Cagliari, Italy\\
$^{23}$ Dipartimento di Fisica dell'Universit\`{a} and Sezione INFN, Trieste, Italy\\
$^{24}$ Dipartimento di Fisica dell'Universit\`{a} and Sezione INFN, Turin, Italy\\
$^{25}$ Dipartimento di Fisica e Astronomia dell'Universit\`{a} and Sezione INFN, Bologna, Italy\\
$^{26}$ Dipartimento di Fisica e Astronomia dell'Universit\`{a} and Sezione INFN, Catania, Italy\\
$^{27}$ Dipartimento di Fisica e Astronomia dell'Universit\`{a} and Sezione INFN, Padova, Italy\\
$^{28}$ Dipartimento di Fisica `E.R.~Caianiello' dell'Universit\`{a} and Gruppo Collegato INFN, Salerno, Italy\\
$^{29}$ Dipartimento DISAT del Politecnico and Sezione INFN, Turin, Italy\\
$^{30}$ Dipartimento di Scienze MIFT, Universit\`{a} di Messina, Messina, Italy\\
$^{31}$ Dipartimento Interateneo di Fisica `M.~Merlin' and Sezione INFN, Bari, Italy\\
$^{32}$ European Organization for Nuclear Research (CERN), Geneva, Switzerland\\
$^{33}$ Faculty of Electrical Engineering, Mechanical Engineering and Naval Architecture, University of Split, Split, Croatia\\
$^{34}$ Faculty of Engineering and Science, Western Norway University of Applied Sciences, Bergen, Norway\\
$^{35}$ Faculty of Nuclear Sciences and Physical Engineering, Czech Technical University in Prague, Prague, Czech Republic\\
$^{36}$ Faculty of Physics, Sofia University, Sofia, Bulgaria\\
$^{37}$ Faculty of Science, P.J.~\v{S}af\'{a}rik University, Ko\v{s}ice, Slovak Republic\\
$^{38}$ Frankfurt Institute for Advanced Studies, Johann Wolfgang Goethe-Universit\"{a}t Frankfurt, Frankfurt, Germany\\
$^{39}$ Fudan University, Shanghai, China\\
$^{40}$ Gangneung-Wonju National University, Gangneung, Republic of Korea\\
$^{41}$ Gauhati University, Department of Physics, Guwahati, India\\
$^{42}$ Helmholtz-Institut f\"{u}r Strahlen- und Kernphysik, Rheinische Friedrich-Wilhelms-Universit\"{a}t Bonn, Bonn, Germany\\
$^{43}$ Helsinki Institute of Physics (HIP), Helsinki, Finland\\
$^{44}$ High Energy Physics Group,  Universidad Aut\'{o}noma de Puebla, Puebla, Mexico\\
$^{45}$ Horia Hulubei National Institute of Physics and Nuclear Engineering, Bucharest, Romania\\
$^{46}$ Indian Institute of Technology Bombay (IIT), Mumbai, India\\
$^{47}$ Indian Institute of Technology Indore, Indore, India\\
$^{48}$ INFN, Laboratori Nazionali di Frascati, Frascati, Italy\\
$^{49}$ INFN, Sezione di Bari, Bari, Italy\\
$^{50}$ INFN, Sezione di Bologna, Bologna, Italy\\
$^{51}$ INFN, Sezione di Cagliari, Cagliari, Italy\\
$^{52}$ INFN, Sezione di Catania, Catania, Italy\\
$^{53}$ INFN, Sezione di Padova, Padova, Italy\\
$^{54}$ INFN, Sezione di Pavia, Pavia, Italy\\
$^{55}$ INFN, Sezione di Torino, Turin, Italy\\
$^{56}$ INFN, Sezione di Trieste, Trieste, Italy\\
$^{57}$ Inha University, Incheon, Republic of Korea\\
$^{58}$ Institute for Gravitational and Subatomic Physics (GRASP), Utrecht University/Nikhef, Utrecht, Netherlands\\
$^{59}$ Institute of Experimental Physics, Slovak Academy of Sciences, Ko\v{s}ice, Slovak Republic\\
$^{60}$ Institute of Physics, Homi Bhabha National Institute, Bhubaneswar, India\\
$^{61}$ Institute of Physics of the Czech Academy of Sciences, Prague, Czech Republic\\
$^{62}$ Institute of Space Science (ISS), Bucharest, Romania\\
$^{63}$ Institut f\"{u}r Kernphysik, Johann Wolfgang Goethe-Universit\"{a}t Frankfurt, Frankfurt, Germany\\
$^{64}$ Instituto de Ciencias Nucleares, Universidad Nacional Aut\'{o}noma de M\'{e}xico, Mexico City, Mexico\\
$^{65}$ Instituto de F\'{i}sica, Universidade Federal do Rio Grande do Sul (UFRGS), Porto Alegre, Brazil\\
$^{66}$ Instituto de F\'{\i}sica, Universidad Nacional Aut\'{o}noma de M\'{e}xico, Mexico City, Mexico\\
$^{67}$ iThemba LABS, National Research Foundation, Somerset West, South Africa\\
$^{68}$ Jeonbuk National University, Jeonju, Republic of Korea\\
$^{69}$ Johann-Wolfgang-Goethe Universit\"{a}t Frankfurt Institut f\"{u}r Informatik, Fachbereich Informatik und Mathematik, Frankfurt, Germany\\
$^{70}$ Korea Institute of Science and Technology Information, Daejeon, Republic of Korea\\
$^{71}$ KTO Karatay University, Konya, Turkey\\
$^{72}$ Laboratoire de Physique des 2 Infinis, Ir\`{e}ne Joliot-Curie, Orsay, France\\
$^{73}$ Laboratoire de Physique Subatomique et de Cosmologie, Universit\'{e} Grenoble-Alpes, CNRS-IN2P3, Grenoble, France\\
$^{74}$ Lawrence Berkeley National Laboratory, Berkeley, California, United States\\
$^{75}$ Lund University Department of Physics, Division of Particle Physics, Lund, Sweden\\
$^{76}$ Nagasaki Institute of Applied Science, Nagasaki, Japan\\
$^{77}$ Nara Women{'}s University (NWU), Nara, Japan\\
$^{78}$ National and Kapodistrian University of Athens, School of Science, Department of Physics , Athens, Greece\\
$^{79}$ National Centre for Nuclear Research, Warsaw, Poland\\
$^{80}$ National Institute of Science Education and Research, Homi Bhabha National Institute, Jatni, India\\
$^{81}$ National Nuclear Research Center, Baku, Azerbaijan\\
$^{82}$ National Research and Innovation Agency - BRIN, Jakarta, Indonesia\\
$^{83}$ Niels Bohr Institute, University of Copenhagen, Copenhagen, Denmark\\
$^{84}$ Nikhef, National institute for subatomic physics, Amsterdam, Netherlands\\
$^{85}$ Nuclear Physics Group, STFC Daresbury Laboratory, Daresbury, United Kingdom\\
$^{86}$ Nuclear Physics Institute of the Czech Academy of Sciences, Husinec-\v{R}e\v{z}, Czech Republic\\
$^{87}$ Oak Ridge National Laboratory, Oak Ridge, Tennessee, United States\\
$^{88}$ Ohio State University, Columbus, Ohio, United States\\
$^{89}$ Physics department, Faculty of science, University of Zagreb, Zagreb, Croatia\\
$^{90}$ Physics Department, Panjab University, Chandigarh, India\\
$^{91}$ Physics Department, University of Jammu, Jammu, India\\
$^{92}$ Physics Program and International Institute for Sustainability with Knotted Chiral Meta Matter (SKCM2), Hiroshima University, Hiroshima, Japan\\
$^{93}$ Physikalisches Institut, Eberhard-Karls-Universit\"{a}t T\"{u}bingen, T\"{u}bingen, Germany\\
$^{94}$ Physikalisches Institut, Ruprecht-Karls-Universit\"{a}t Heidelberg, Heidelberg, Germany\\
$^{95}$ Physik Department, Technische Universit\"{a}t M\"{u}nchen, Munich, Germany\\
$^{96}$ Politecnico di Bari and Sezione INFN, Bari, Italy\\
$^{97}$ Research Division and ExtreMe Matter Institute EMMI, GSI Helmholtzzentrum f\"ur Schwerionenforschung GmbH, Darmstadt, Germany\\
$^{98}$ Saga University, Saga, Japan\\
$^{99}$ Saha Institute of Nuclear Physics, Homi Bhabha National Institute, Kolkata, India\\
$^{100}$ School of Physics and Astronomy, University of Birmingham, Birmingham, United Kingdom\\
$^{101}$ Secci\'{o}n F\'{\i}sica, Departamento de Ciencias, Pontificia Universidad Cat\'{o}lica del Per\'{u}, Lima, Peru\\
$^{102}$ Stefan Meyer Institut f\"{u}r Subatomare Physik (SMI), Vienna, Austria\\
$^{103}$ SUBATECH, IMT Atlantique, Nantes Universit\'{e}, CNRS-IN2P3, Nantes, France\\
$^{104}$ Sungkyunkwan University, Suwon City, Republic of Korea\\
$^{105}$ Suranaree University of Technology, Nakhon Ratchasima, Thailand\\
$^{106}$ Technical University of Ko\v{s}ice, Ko\v{s}ice, Slovak Republic\\
$^{107}$ The Henryk Niewodniczanski Institute of Nuclear Physics, Polish Academy of Sciences, Cracow, Poland\\
$^{108}$ The University of Texas at Austin, Austin, Texas, United States\\
$^{109}$ Universidad Aut\'{o}noma de Sinaloa, Culiac\'{a}n, Mexico\\
$^{110}$ Universidade de S\~{a}o Paulo (USP), S\~{a}o Paulo, Brazil\\
$^{111}$ Universidade Estadual de Campinas (UNICAMP), Campinas, Brazil\\
$^{112}$ Universidade Federal do ABC, Santo Andre, Brazil\\
$^{113}$ University of Cape Town, Cape Town, South Africa\\
$^{114}$ University of Houston, Houston, Texas, United States\\
$^{115}$ University of Jyv\"{a}skyl\"{a}, Jyv\"{a}skyl\"{a}, Finland\\
$^{116}$ University of Kansas, Lawrence, Kansas, United States\\
$^{117}$ University of Liverpool, Liverpool, United Kingdom\\
$^{118}$ University of Science and Technology of China, Hefei, China\\
$^{119}$ University of South-Eastern Norway, Kongsberg, Norway\\
$^{120}$ University of Tennessee, Knoxville, Tennessee, United States\\
$^{121}$ University of the Witwatersrand, Johannesburg, South Africa\\
$^{122}$ University of Tokyo, Tokyo, Japan\\
$^{123}$ University of Tsukuba, Tsukuba, Japan\\
$^{124}$ University Politehnica of Bucharest, Bucharest, Romania\\
$^{125}$ Universit\'{e} Clermont Auvergne, CNRS/IN2P3, LPC, Clermont-Ferrand, France\\
$^{126}$ Universit\'{e} de Lyon, CNRS/IN2P3, Institut de Physique des 2 Infinis de Lyon, Lyon, France\\
$^{127}$ Universit\'{e} de Strasbourg, CNRS, IPHC UMR 7178, F-67000 Strasbourg, France, Strasbourg, France\\
$^{128}$ Universit\'{e} Paris-Saclay Centre d'Etudes de Saclay (CEA), IRFU, D\'{e}partment de Physique Nucl\'{e}aire (DPhN), Saclay, France\\
$^{129}$ Universit\`{a} degli Studi di Foggia, Foggia, Italy\\
$^{130}$ Universit\`{a} del Piemonte Orientale, Vercelli, Italy\\
$^{131}$ Universit\`{a} di Brescia, Brescia, Italy\\
$^{132}$ Variable Energy Cyclotron Centre, Homi Bhabha National Institute, Kolkata, India\\
$^{133}$ Warsaw University of Technology, Warsaw, Poland\\
$^{134}$ Wayne State University, Detroit, Michigan, United States\\
$^{135}$ Westf\"{a}lische Wilhelms-Universit\"{a}t M\"{u}nster, Institut f\"{u}r Kernphysik, M\"{u}nster, Germany\\
$^{136}$ Wigner Research Centre for Physics, Budapest, Hungary\\
$^{137}$ Yale University, New Haven, Connecticut, United States\\
$^{138}$ Yonsei University, Seoul, Republic of Korea\\
$^{139}$  Zentrum  f\"{u}r Technologie und Transfer (ZTT), Worms, Germany\\
$^{140}$ Affiliated with an institute covered by a cooperation agreement with CERN\\
$^{141}$ Affiliated with an international laboratory covered by a cooperation agreement with CERN.\\

\end{flushleft} 
  
\end{document}